\documentclass[aps,pra,twocolumn,amsmath,amssymb,nofootinbib,showpacs,superscriptaddress]{revtex4-1}
\usepackage[english]{babel}
\usepackage{latexsym}
\usepackage{graphics}
\usepackage{graphicx}
\usepackage{epsfig}
\usepackage{color}
\usepackage{bm}
\usepackage{amsmath}
\usepackage{amssymb}
\usepackage{amsthm}
\usepackage{dcolumn}
\usepackage{bm}
\usepackage{float}
\usepackage{hyperref}
\usepackage{color}
\usepackage{epstopdf}
\usepackage{braket}
\usepackage{cleveref}
\usepackage[svgnames]{xcolor}
\hypersetup{hidelinks,colorlinks=true,allcolors=DarkBlue}

\theoremstyle{remark}

\begin{document}

\preprint{APS/123-QED}

\title{Improving the performance of fermionic neural networks \\ with the Slater exponential Ansatz}

\author{Denis Bokhan}
\email{denisbokhan@mail.ru}
\affiliation{Laboratory of Molecular Beams, Physical Chemistry Division, Department of Chemistry, Lomonosov Moscow State University, Moscow 119991, Russia}
\affiliation{Russian Quantum Center, Skolkovo, Moscow 143025, Russia}

\author{Maria M. Kolchenko}
\affiliation{Laboratory of Molecular Beams, Physical Chemistry Division, Department of Chemistry, Lomonosov Moscow State University, Moscow 119991, Russia}
\affiliation{Russian Quantum Center, Skolkovo, Moscow 143025, Russia}

\author{Aleksey S. Boev}
\affiliation{Russian Quantum Center, Skolkovo, Moscow 143025, Russia}
\affiliation{National University of Science and Technology ``MISIS'', Moscow 119049, Russia}

\author{Aleksey K. Fedorov}
\email{akf@rqc.ru}
\affiliation{Russian Quantum Center, Skolkovo, Moscow 143025, Russia}
\affiliation{National University of Science and Technology ``MISIS'', Moscow 119049, Russia}

\author{Dmitrii N. Trubnikov}
\affiliation{Laboratory of Molecular Beams, Physical Chemistry Division, Department of Chemistry, Lomonosov Moscow State University, Moscow 119991, Russia}

\date{\today}
\begin{abstract}
In this work, we propose a technique for the use of fermionic neural networks (FermiNets) with the Slater exponential Ansatz for electron-nuclear and electron-electron distances, 
which provides faster convergence of target ground-state energies due to a better description of the interparticle interaction in the vicinities of the coalescence points. 
Our analysis of learning curves indicates on the possibility to obtain accurate energies with smaller batch sizes using arguments of the bagging approach. 
In order to obtain even more accurate results for the ground-state energies, we propose an extrapolation scheme for estimating Monte Carlo integrals in the limit of an infinite number of points. 
Numerical tests for a set of molecules demonstrate a good agreement with the results of the original FermiNets approach (achieved with larger batch sizes than required by our approach) 
and with results of the coupled-cluster singles and doubles with perturbative triples (CCSD(T)) method that are calculated in the complete basis set (CBS) limit.
\end{abstract}

\maketitle

\section{Introduction}

Description of complex many-body quantum systems is a tremendous challenge, which is of primary importance for fundamental science as well as practical problems in physics, chemistry, biology, and materials science~\cite{Bravyi2020}.
In the quantum chemistry domain, the amount of required resources  for ab-initio calculations of properties of molecules grows exponentially with the size of the system. 
Methods based on the coupled-cluster (CC) approach (for a review, see Ref.~\cite{Bartlett2007}) are successfully applied to a wide range of quantum chemical problems. 
However, the computational cost of the CC scheme is still prohibitively high for its application to large-scale systems. 
Particularly, the coupled-cluster singles and doubles with perturbative triples (CCSD(T)) method~\cite{Raghavachari1989}, which is considered as a ``golden standard" in quantum chemical calculations, 
scales as $N^{7}$ with the number of basis functions $N$. 
On the other hand, in order to calculate thermochemical properties, such as reaction enthalpies or atomization energies with chemical accuracy (1 kcal/mol), atomic orbital bases with high angular momentum are required, 
which makes such calculations to be of extremely high computational costs.
The issue can be partially solved via utilization of the explicitly-correlated coupled cluster method (CC-F12)~\cite{Noga1994},
where it is possible to use atomic orbitals bases with reduced maximal angular momentum. 
Approximated versions of the CC-F12 approach, such as CCSD(T)(F12)~\cite{Tew2008,Bokhan2009}, CCSD(T)-F12x~\cite{Knizia2009}, and CCSD(F12*)(T)~\cite{Hattig2010}, 
became a commonly used tool for highly-accurate calculations of molecular properties. 
These methods, however, still have $N^{7}$ scaling and are not directly applicable for large-scale systems. 
Another widely used approach of quantum chemistry is the density functional theory (DFT)~\cite{Kohn1965} and large variety of functionals is available nowadays for practical applications~\cite{Jones2015}. 
Success of the DFT method is related to its low computational cost of $N^{3}$ scaling with the size of used basis. 
On the other hand, DFT methods have a limited accuracy and mostly used in the cases with low precision requirements.
A variety of effects including practically relevant problems of physics and chemistry are thus beyond the scope of DFT~\cite{Troyer2016}.
Serious attention is paid to the idea of using quantum computing devices for solving quantum chemistry problems (for a review, see Refs.~\cite{Bravyi2020,Aspuru-Guzik2019,Aspuru-Guzik2020,Fedorov2021}),
however the power of this approach is mainly limited by the capabilities of existing quantum hardware. 

Impractical scaling of existing \textit{ab initio} methods essentially raises a problem of finding more compact representations of wave functions of chemical systems. 
A point-wise representation on the numerical grid without the use of any basis of atomic orbitals can be an attractive alternative. 
This is because the number of points scales linearly with the number of similar size atoms in the system of interest. 
For such kind of representations, variational quantum Monte Carlo (VQMC) is a typical strategy for calculation of ground state energies.
More advanced projector methods, such as diffusion quantum Monte Carlo (DQMC)~\cite{Foulkes2001} and auxiliary-field quantum Monte Carlo (AFQMC)~\cite{Zhang2018}, have been developed. 
Most common \textit{Ans\"atze} for continuous-space many-electron problems in three dimensions (3D) that used in VQMC are Slater-Jastrow and Slater-Jastrow backflow wave functions~\cite{Foulkes2001,Ceperley1978,Feynman1956}. 
The latter performs an additional transformation to the orbitals (before their evaluation) by shifting the position of each electron by an amount dependent on the positions of nearby electrons. 
Alternative forms of wave functions, such as the Pfaffian functional form~\cite{Bajdich2006} and tensor networks~\cite{Orus2014} for VQMC method, have been intensively studied. 
Nevertheless, despite the high accuracy of VQMC Slater-Jastrow methods, currently their practical application is limited to rather a small systems.

In recent years, the representation of wave functions using neural networks has emerged as powerful tool for studying properties of quantum many-body systems~\cite{Troyer2017}.
This gives rise to a family of \textit{Ans\"atze}, known as neural network quantum states (NNQS)~\cite{Troyer2017,Carleo2019,Melko2019}. 
Various types of neural networks have been considered as the basis for such representation, specifically, a significant attention has been paid to Restricted Boltzman Machines (RBMs)~\cite{Troyer2017}.
Although NNQS \textit{Ans\"atze} have been successfully used to investigate various problems in condensed matter physics, their direct applications to fermionic problems, and particularly quantum chemistry~\cite{Xia2018}, have been very limited until recently. 
Recent advanced are related, first, to pushing the capabilities of the RBM-based approach~\cite{Carleo2020} and to the use of an autoregressive neural network model for treatment of electronic correlation problem~\cite{Lvovsky2022},
which approaches FCI-quality solutions on systems with up to 30 spin orbitals.
An alternative path to overcome existing limitations is to use neural networks to encode real-space wave functions with flexible basis sets.
Recent successes in this directions are related Fermionic Neural Network (FermiNet)~\cite{Pfau2020} and Pauli Neural Network (PauliNet)~\cite{Hermann2020} approaches.

Within neural-network-based approach, each step of calculations consists of two parts: (i) training of neural networks and (ii) subsequent Monte Carlo integration. 
The batch of electronic coordinates is required to construct input features for neural networks and the same batch is used for Monte Carlo integration. 
Sampling the random electronic coordinates should be done from the distribution, which corresponds to the electron density, and, thus, 
Markov-Chain Monte Carlo (MCMC) can be utilized for the evaluation of energy functional with neural network wave functions. 
Typically, a random walk algorithm is used for this purpose. 
In order to obtain more accurate results for target ground state energies, large batch sizes are required. 
In particular, previously reported results of FermiNet calculations even with batches of size 4096 points~\cite{Pfau2020} were not always accurate due to the errors in the Monte Carlo integration. 
Then for a wider adoption of neural-network-based approach it becomes crucial to develop a strategy for calculating the target VQMC ground state energies with smaller batch sizes without significant loss in accuracy.

In this work, we present a modification of input features of FermiNets by introducing the Slater exponential \textit{Ansatz} for electron-nuclear and electron-electron distances.
We develop an extrapolation method, which is able to provide accurate ground-state energies with reduced numbers of grid points giving the way to train FermiNets with smaller batch sizes.
Our numerical results on the basis of the test set of molecules show the possibility to improve FermiNet approach for practically relevant chemical problems.

Our work is organized as follows.
In Sec.~\ref{sec:theory}, we provide the general theory behind the FermiNet approach with exponential \textit{Ansatz} inputs.
In Sec.~\ref{sec:results}, we demonstrates results of calculations of molecular energies and reaction enthalpies.
Our comparison with the standard FermiNet method indicate that the proposed Ansatz gives the way to train FermiNets with significantly smaller batch size without loss of accuracy.
In Sec.~\ref{sec:reaction}, we consider reaction enthalpies.
We summarize our results in Sec.~\ref{sec:conlcusion}.

\section{Fermionic Neural Networks with exponential \textit{Ansatz} inputs}\label{sec:theory}

\subsection{VQMC method}

Let us we briefly remind the basics of the VQMC method.
As in the standard variational method energy functional,
\begin{equation}
	L(\Theta) = \frac{\langle\Psi(\Theta)|\hat{H}|\Psi(\Theta)\rangle}{\langle\Psi(\Theta)|\Psi(\Theta)\rangle},
\end{equation}
is minimized with respect to parameters $\Theta$ in order to find the ground-state energy:
\begin{equation}
	E =\arg\min_{\Theta}\left[L(\Theta)\right].
\end{equation}
The density $\rho(\Theta)$ and the local energy $E_{L}(x,\Theta)$ have following form:
\begin{equation}
	\rho(\Theta) = \frac{|\Psi(\Theta)|^{2}}{\langle\Psi(\Theta)|\Psi(\Theta)\rangle},
\end{equation}
\begin{equation}
	E_{L}(x,\Theta) = \frac{\hat{H}\Psi(x,\Theta)}{\Psi(x,\Theta)},
\end{equation}
where $x$ stands for all electronic coordinates. 
In the explicit form, $E_{L}(x,\Theta)$ can be presented as follows:
\begin{equation}
	E_{L} = \frac{(T+V)\Psi(x,\Theta)}{\Psi(x,\Theta)} = \frac{T\Psi(x,\Theta)}{\Psi(x,\Theta)} + V(x)
\end{equation}
While potential energy operator $V(x)$ gives trivial contribution of itself, the term with kinetic energy operator $T$ can be written as:
\begin{equation}
	\frac{T\Psi(x,\Theta)}{\Psi(x,\Theta)} = -\frac{1}{2}\sum_{i=1}^{N}
	\left[\left(\frac{\partial \ln |\Psi|}{\partial r_{i}}\right)^{2}
	+ \frac{\partial^{2} \ln |\Psi|}{\partial r_{i}}^{2}
	\right]
\end{equation}
where $N$ is the number of electrons and $r_{i} = x_{i},y_{i},z_{i}$ for $i$th electron.
In terms of $\rho(\Theta)$ and $E_{L}(x,\Theta)$, the energy functional can be rewritten as the expectation value with distribution function, corresponding to the density
\begin{equation}{\label{eqn:EXP}}
	L(\Theta) = \int \rho (x, \Theta) E_{L} (x,\Theta) dx = \mathbb{E}_{\rho (x,\Theta) }( E_{L} (x,\Theta)).
\end{equation}
Sampling coordinates $x$ is repeated until $L(\Theta)$ is converged to a certain predefined threshold of the accuracy. 

\subsection{Extrapolation procedure for QMC energies}\label{sec:Extrapolation}

In order to develop the extrapolation procedure towards infinite number of points, we need an expression for the error estimation of the certain $N$-dimensional Monte Carlo integral $I$, 
which is calculated with $N$ grid points:
\begin{equation}
	I_{\rm exact} = I_{N} \pm V\frac{\sigma_{N}}{\sqrt{N}},
\end{equation}
where $\sigma_{N}$ is the variance of integrand and
\begin{equation}
	V = \int_{V}d\mathbf{x}
\end{equation}
is the volume of system.
We note that the error grows linearly with the size of the system, and, thus, large errors can be expected for systems with many electrons even for large batch sizes. 
For two number of integration points $N_{1}$ and $N_{2}$, we can write:
\begin{equation}
	I_{\rm exact} = I_{N_{1}} \pm V\frac{\sigma_{N_{1}}}{\sqrt{N_{1}}}, 
	\quad
	I_{\rm exact} = I_{N_{2}} \pm V\frac{\sigma_{N_{2}}}{\sqrt{N_{2}}}.
\end{equation}
If we take large enough number of points $N_{1}$ and $N_{2}$, one can introduce an approximation $\sigma_{N_{1}} \approx \sigma_{N_{2}}$, 
so the above equations can be rewritten as:
\begin{equation}{\label{eqn:E1}}
	I_{\rm exact} = I_{N_{1}} \pm \frac{\rm const}{\sqrt{N_{1}}},
	\quad
	I_{\rm exact} = I_{N_{2}} \pm \frac{\rm const}{\sqrt{N_{2}}},
\end{equation}
where ${\rm const} = V \sigma_{N_{1}}$. 
Now we consider the case, when for the case of $N_{2} > N_{1}$ obtained integral become more negative:
\begin{equation}{\label{eqn:E1_2}}
        I_{N_{2}} < I_{N_{1}}.
\end{equation}
(In fact, for all molecular systems, considered below in Sec.~\ref{sec:results} this is always the case). 
If we further assume the monotonic behaviour of $I_{N}$ with the increase of number of integration points, 
we expect that value of $I_{\rm exact}$ becomes more negative then $I_{N_{1}}$. 
In other words,
\begin{equation}{\label{eqn:E1_3}}
	I_{\rm exact} = I_{N_{1}} - \frac{\rm const}{\sqrt{N_{1}}}.
\end{equation}
For the case of $I_{N_{2}}$ we still have
\begin{equation}{\label{eqn:E1_4}}
	I_{\rm exact} = I_{N_{2}} \pm \frac{\rm const}{\sqrt{N_{2}}}.
\end{equation}
For the negative sign in the right-hand side (rhs) of Eq.~(\ref{eqn:E1_4}) combination of it with Eq.~(\ref{eqn:E1_3}) gives
\begin{equation}
	I_{exact}^{L} = \frac{ I_{N_{2}}\sqrt{N_{2}} - I_{N_{1}}\sqrt{N_{1}}}{\sqrt{N_{2}}-\sqrt{N_{1}}}.
\end{equation}
Similarly, for the case of plus sign in rhs of Eq.~(\ref{eqn:E1_4}) result of combination with Eq.~(\ref{eqn:E1_3}) can be written as:
\begin{equation}
	I_{eaxact}^{R} = \frac{ I_{N_{2}}\sqrt{N_{2}} + I_{N_{1}}\sqrt{N_{1}}}{\sqrt{N_{2}}+\sqrt{N_{1}}}.
\end{equation}
Values of $I_{exact}^{L}$ and $I_{exact}^{R}$ are obtained from Eq.~(\ref{eqn:E1_4}) and Eq.~(\ref{eqn:E1_3}), 
which represent estimation of maximal possible deviations of $I_{exact}$ from calculated integrals $I_{N}$. 
On the other hand, maximal errors occur with smallest probability, what means that $I_{exact}^{L}$ and $I_{exact}^{R}$ correspond to the least probable values of
exact integral. In our scheme the most probable value of target integral is assumed to be an average of its least probable values:
\begin{equation}
        I_{exact} = \frac{I_{exact}^{L} + I_{exact}^{R}}{2}.
\end{equation}
This assumption in similar to the case of Gaussian distribution with least probable values on its left and right limits, while
most probable value lays in the middle of those two least probable values. 
We note that extrapolated value does not depend upon volume of the system, and, thus we can expect that extrapolation  error will grow rather modestly with the increase of system size.

\begin{widetext}

\subsection{Exponential \textit{Ansatz} for FermiNets}

The use of FermiNet \textit{Ansatz} in the context of VQMC approach implies target wave functions of the following form:
\begin{equation}{\label{eqn:FN}}
	\Psi(r_{1}^{\uparrow}...r_{N/2}^{\uparrow},r_{1}^{\downarrow}...r_{N/2}^{\downarrow}) = \sum_{k=1}^{N_{\rm det}}
	C_{k} \det[\varphi_{i}^{k\uparrow}(r_{j}^{k\uparrow}; \{ r_{/j}^{k\uparrow} \}; \{ r^{k\downarrow}  \} )]
	\det[\varphi_{i}^{k\downarrow}(r_{j}^{k\downarrow}; \{ r_{/j}^{k\downarrow} \}; \{ r^{k\uparrow}  \} )],
\end{equation}
where $N_{\rm det}$ is the number of determinants (${\rm det}$) that imply the construction of corresponding Slater determinant and $r_{/j}^{k\uparrow}$ means coordinate vectors of all electrons except $j$th one. 
Unlike regular definition of orbitals in traditional quantum chemistry, here we have orbitals, which are not single-particle functions; each 
$\varphi_{i}^{k\uparrow}(r_{i}^{k\uparrow}; \{ r_{/i}^{k\uparrow} \}; \{ r^{k\downarrow}  \} )$ 
depends upon coordinates of all other electrons. 
Such definition of orbitals resembles Jastrow-Slater backflow \textit{Ansatz}:
\begin{equation}{\label{eqn:Jast}}
	\Psi(r_{1}^{\uparrow}...r_{N/2}^{\uparrow},r_{1}^{\downarrow}...r_{N/2}^{\downarrow})^{JSB} = \sum_{k=1}^{N_{\rm det}}
	C_{k} \det[\varphi_{i}^{k\uparrow}(q_{j}^{k\uparrow})]
	\det[\varphi_{i}^{k\downarrow}(q_{j}^{k\downarrow})],
	e^{J(r,R)}
\end{equation}
where $\exp(J(r,R))$ is the Jastrow factor and $q_{i} = r_{i} + \xi_{i}(\{ r_{/i} \})$ is the generalized electronic coordinate after transformation with backflow functions $\xi$. 
The role of Jastrow factors in Eq.~(\ref{eqn:Jast}) is to describe behavior of target wave functions in the vicinities of interparticle coalescence points. 

We would like to note that there are no Jastrow factors in Eq.~(\ref{eqn:FN}). 
Instead, FermiNet takes as inputs sets of electron-nuclear distance vectors $r_{i}-R_{\alpha}$ and the corresponding distance values $|r_{i}-R_{\alpha}|$ for each electron $i$ and nucleus $\alpha$. 
Similarly two-particle streams of FermiNet take $r_{i}-r_{j}$ and $|r_{i}-r_{j}|$ as input features. 
The use of $|r_{i}-R_{\alpha}|$ and $|r_{i}-r_{j}|$ is required for the description of short-range correlation effects within FermiNet wave functions of Eq.~(\ref{eqn:FN}).
Such set of input features would be enough at short distances between electrons or between electron and nucleus, 
since at the vicinities of coalescence points of electron $i$ and nucleus $\alpha$ wave function can be written as follows:
\begin{equation}
	\Psi(r_{1}...r_{N},R_{1}..R_{M}) \propto (1 + C_{1}|r_{i}-R_{\alpha}|+O(|r_{i}-R_{\alpha}|^{2})),
\end{equation}
where the constant $C_{1}$ is proportional to the charge of the nucleus. 
In the vicinity of the coalescence of electrons $i$ and $j$, the wave function has following form:
\begin{equation}
	\Psi(r_{1}...r_{N},R_{1}..R_{M}) \propto (1 + C_{2}|r_{i}-r_{j}|+O(|r_{i}-r_{j}|^{2})),
\end{equation}
where $C_{2}$ equal to ${1}/{2}$ or ${1}/{4}$ for singlet or triplet pairs, correspondingly.
At the large distances between particle correlation factors should naturally decay, as it happens for $\exp(J(r,R))$, and, thus, we can expect a similar behavior from short-range correlation input features of Ferminet. 
In contrast, $|r_{i}-R_{\alpha}|$ and $|r_{i}-r_{j}|$ does not vanish at large distances and that may bring additional difficulties in training the neural network. 
\end{widetext}

In order to overcome this problem, we propose to use Slater exponential short-range correlation features:
\begin{equation}{\label{eqn:FEN}}
	f_{i\alpha} = \frac{1}{\beta_{\alpha}}(1-\exp(-\beta_{\alpha}|r_{i}-R_{\alpha}|)),
\end{equation}
\begin{equation}{\label{eqn:F12}}
	f_{ij} = \frac{1}{\gamma_{ij}}(1-\exp(-\gamma_{ij}|r_{i}-r_{j}|)).
\end{equation}
Slater exponential correlation factors~\cite{Ten2004} are widely used in the modern explicitly-correlated quantum chemistry calculations. 
The idea of such correlation factors has been proposed for MP2-F12 method in Refs.~\cite{Ten2014-2,Ten2007}. 
At short distances, features in Eqs.~(\ref{eqn:FEN}) and (\ref{eqn:F12}) coincide with the original $|r_{i}-R_{\alpha}|$ and $|r_{i}-r_{j}|$ features; 
division by $\beta_{\alpha}$ and $\gamma_{ij}$ of $(1-\exp(-\beta_{\alpha}|r_{i}-R_{\alpha}|))$ and $(1-\exp(-\gamma_{ij}|r_{i}-r_{j}|))$ is needed for exact satisfaction of cusp conditions. 

Slater-type correlation factors are more suitable for simulation of Coulomb hole then the linear factors. 
Utilization of former brings significant increase in accuracy of target correlation energies. Modified FermiNet algorithm with newly introduced features can be presented by pseudocode (see Appendix A). 
The use of Conv1D and Conv2D functions within convolutional neural network layers guarantees the permutational equivariance of FermiNet model with respect to electronic coordinates. 
Such equivariance ensures that FermiNet orbitals $\varphi_{i}^{k\uparrow}(r_{j}^{k\uparrow}; \{ r_{/j}^{k\uparrow} \}; \{ r^{k\downarrow}  \} )$ remain unchanged after permutations of orbital indices within $\{ r_{/j}^{k\uparrow} \}$ and $\{ r^{k\downarrow}  \}$ sets. 
We note that the parameter $\beta_{\alpha}$ in Eq.~(\ref{eqn:FEN}) does not depend upon the electron index $i$, i.e. around some nucleus $\alpha$ all electrons are described by the same short-range part of wave function. 
In fact, this not only simplifies the model, but also preserves permutational equivariance of modified FermiNet model. 
In the case of closed-shell systems new input features of Eqs.~(\ref{eqn:FEN}) and (\ref{eqn:F12}) introduce two sets of additional parameters: 
the number of $\beta$ parameters is equal to the number of atoms, while number of $\gamma$ parameters is equal to $(N_{el}/2)(N_{el}/2-1)/2$, 
which is typically much smaller than overall number of FermiNet trainable variables.

\section{Calculations of molecular energies using the FermiNet with the exponential \textit{Ansatz}}\label{sec:results}

In order to estimate the accuracy of the developed procedure we calculate ground-state energies of a set of molecules and compare it with original FermiNet results presented in Ref.~\cite{Pfau2020}. 
All the considered molecules are calculated in its equilibrium geometries. 
We use the energies available at CCSD(T) level in CBS limit as an etalon for  comparison.
All the results were obtained using 16 Slater determinants in the expansion of FermiNet wave functions of Eq.~(\ref{eqn:FN}).

As it is done in Ref ~\cite{Pfau2020}, we use 4-layer FermiNet with 256 hidden units for the one-electron stream and 32 hidden units for the two electron stream.
A tanh-type nonlinearity is employed for all network layers. 
Full anisotropic envelopes are utilized to ensure correct asymptotics of obtained orbitals. 
For the implementation of new inputs features, we use both TensorFlow and Jax FermiNet codes, which are available online~\cite{FermiNet}. Our setup for calculations include  1 GPU 
Nvidia Tesla T4 with 16GB of memory for molecules up to N$_{2}$ and CO, while for C$_{2}$H$_{4}$ and C$_{4}$H$_{6}$ Nvidia Quadro RTX5000 with 16GB of memory is utilized. Computational time, which has been spent for test calculation, varies from several hours for small molecules like LiH to 10 days for C$_{4}$H$_{6}$ molecule, which is calculated using 4 GPUs.
Results of numerical experiments are presented in Table~\ref{bra:tab1}.

The reduced dependence of the accuracy on the batch size has enabled achieving reasonably accurate values of target ground-state energies with significantly smaller number of points than 4096 as in original FermiNet experiments. 
Specifically, in the case of using 250 points for H atom and 500 points for C, N, O, and F atoms it is possible to get highly accurate results for small systems, such as LiH, Li$_{2}$, CH$_{2}$, and HF molecules. 
Within the framework of the QMC approach, energies are calculated as converged mean values over FermiNet learning curves and this fact gives us a possibility to look at the results from the point of view of the random forest algorithm, 
which is typically used for the case of many weak predictors. 
Since small batch sizes are used, we have errors in energies and gradients, and, thus, each point of learning curve corresponds to the weak predictor. 

Samples of electronic coordinates on each training step are not duplicated, so the probability of sampling the same coordinates twice is very low. 
Also, all the samples are uncorrelated with each other. 
So far we observe that predictor voting via averaging can provide reasonably accurate results. 

The main benefits from introducing new input features of Eqs.~(\ref{eqn:FEN}) and (\ref{eqn:F12}) include acceleration of convergence and reduction of the variance of learning curves. 
Fig.~\ref{fig:Fig2} and Fig.~\ref{fig:Fig3} present the learning process for LiH and Li$_{2}$ molecules, correspondingly. 
One can see that even for such small systems the use of the exponential \textit{Ansatz} leads to the significant reduction of number of iterations needed for convergence as well as variance reduction.
Typically it takes around 200,000 iterations to converge with standard input features, while the use of the exponential input features allows one to reduce this number by a factor 2-2.5.
On the other hand, both old and new \textit{Ans\"atze} provide the similar accuracy of obtained ground-state energies. 
We note that neural network is expressive enough to provide correct total energies even when input features with wrong asymptotic behaviour are used.

For larger systems like N$_{2}$, CO, C$_{2}$H$_{4}$, and C$_{4}$H$_{6}$ energies are calculated using extrapolation procedure from Sec.~\ref{sec:Extrapolation}. 
For the case of bicyclobutane the obtained error in ground-state energy in the case of the original FermiNet approach is few tenth of Hartree, while our results have the error around one tenth of a.u. 
Thus, the FermiNet model with improved inputs features and subsequent extrapolation of target ground state energies can be used as a practical tool for the highly-accurate calculations of molecular systems. 

\section{Reaction enthalpies}\label{sec:reaction}

Further estimation of performance of developed approach is done by calculatting enthalpies of 13 isogyric reactions, considered by Bak \textit{et. al.} in Ref.~\cite{Bak1}. 
The obtained results are given in Table \ref{bra:tab2}. 
The contribution of electronic energies to reaction enthalpies are calculated from corresponding experimental data by subtraction of zero point vibration energies (harmonic and unharmonic contributions) and scalar relativistic corrections. 
Molecular equilibrium geometries are taken from Ref.~\cite{Bak2}.
Results of CCSD(T)(F12) method with aug-cc-pCVTZ bases are also used for comparison with our approach; the corresponding values taken from ref. 
In order to analyze accuracy of extrapolation procedure we also provide values of reaction enthalpies, 
which are calculated without extrapolation and in this case values of ground-state energies, obtained with largest batch size ($N_{1}$ or $N_{2}$) are used for calculations. 
The number of points that are used for extrapolation are given for each molecule in Table \ref{bra:tab3}. 
Statistical measures, such as maximal absolute error ($\Delta_{\max}^{\rm abs}$), mean absolute error ($\bar{\Delta}_{\rm abs}$) and standard deviation ($\Delta_{\rm std}$) are used for estimation of performance of all considered methods. 

Both FermiNet method with extrapolation and CCSD(T)(F12) provide results which are close to experimental data. 
In fact, for these two approaches ($\Delta_{\max}^{\rm abs}$) and ($\bar{\Delta}_{\rm abs}$) values does not exceed threshold of chemical accuracy of 4.2 kJ/mol. 
In opposite, FermiNet without extrapolation shows quite poor statistics of errors in reaction enthalpies. 

This fact is not surprising since errors of Monte-Carlo integration grow linearly with the size of the system and for larger molecules, like CO$_{2}$ such errors become very significant. 
Extrapolated FermiNet results show rather a modest increase of errors with the size of system; those errors can be reduced by using larger batch sizes $N_{1}$ and $N_{2}$ if allowed by used hardware. 
In overall, the FermNet approach with the proposed Slater exponential \textit{Ansatz} and extrapolation procedure can be a helpful tool for accurate calculation of molecular energies.

\section{Conclusions}\label{sec:conlcusion}

In this work, we have introduced the extrapolation procedure, which enables us to estimate VQMC integrals in the limit of infinite number of grid points. 
As we have shown, the developed extrapolation technique can provide accurate ground-state energies using reduced numbers of grid points, which gives the possibility to train FermiNets with significantly smaller batch sizes. 
Also, fermionic neural networks were extended by support of new input features that significantly accelerated the training process and reduced overall computational time. 
Such acceleration of the learning procedure is related to the fact that the exponential \textit{Ans\"atze} for interparticle distances are more suitable for description of wave function behavior near the coalescence points. 
We would like point out that the exponential \textit{Ans\"atze} in principle can be used for other types of neural-network-based techniques, for example, potentially in PauliNet~\cite{Hermann2020}.
As we expect the improved neural-network-based approaches can help for predictive calculations of molecular energies for practically relevant quantum chemistry problems.

\section*{Acknowledgments}

A.K.F. acknowledge the support of the Russian Science Foundation (19-71-10092).
The work at the Russian Quantum Center is partially supported by Nissan Research Division, Nissan Motor Co., Ltd., and Nissan Research Center-Russia.
The results of Sec.~\ref{sec:reaction} are obtained with the support of the Priority 2030 program at the National University of Science and Technology ``MISIS'' under the project K1-2022-027.

\section*{Availability of data}
The data that supports the findings of this study are available within the article.


\bibliographystyle{naturemag}
\bibliography{references}

\begin{widetext}

\begin{table}[h]
\begin{center}
\begin{tabular}{|c|c|c|c|c|c|}
\hline
Molecule & \multicolumn{2}{|c|}{Batch size} & $E$ (FermiNet) & $E$ (FermiNet) & $E$ (CCSD(T),CBS) \\
& $N_{1}$& $N_{2}$ & improved & standard & \\ \hline
\hline
LiH & 256 & & $\underline{\mathbf{-8.0707}}$ & -8.0705 & \underline{-8.0707} \\
Li$_{2}$ & 1024 & & $\mathbf{-14.9949}$ & -14.9948 & \underline{-14.9951} \\
CH$_{2}$ & 1000 & & $\underline{\mathbf{-39.1331}}$ & & \underline{-39.1331} \\
HF & 750 & & $\mathbf{-100.4596}$ &  & \underline{-100.4597} \\
N$_{2}$ & 1000 & 1500 & $\mathbf{-109.5430}$ & -109.5388 & \underline{-109.5425} \\
CO & 1000 & 1500 & $\mathbf{-113.3241}$ & -113.3218 & \underline{-113.3255} \\
C$_{2}$H$_{4}$ & 2000 & 2500 & $\mathbf{-78.5910}$ & -78.5844 & \underline{- 78.5888} \\
C$_{4}$H$_{6}$ & 2000 & 2500 & $\mathbf{-155.9471}$ & -155.9263 & \underline{-155.9575}\\ 
\hline
\end{tabular}
\caption{Ground state energies of the set of molecules, which are calculated with standard version of FermiNets and our improvements. Batches of 4096 points are used in calculations with the standard version.
We use bold font to highlight the cases, where our modification of FermiNet outperforms the original FermiNet;
the results matching CCSD(T) and CBS are additionally underlined.}
\label{bra:tab1}
\end{center}
\end{table}

\begin{table}[h]
\begin{center}
\begin{tabular}{|c|c|c|c|c|}
\hline
Reaction & FermiNet & FermiNet & CCSD(T)(F12) & Exp~\cite{Bak1} \\
& improved - & improved & aug-cc-pVTZ & \\
& no extrapolation & & & \\ \hline
\hline
CO + H$_{2}$ $\rightarrow$ CH$_{2}$O & -25.20 & -21.00 & -24.11 & -21.85 \\
N$_{2}$ + 3H$_{2}$ $\rightarrow$ 2NH$_{3}$ & -172.22 & -161.73 & -164.07 & -165.38 \\
C$_{2}$H$_{2}$ + H$_{2}$ $\rightarrow$ C$_{2}$H$_{4}$ & -201.38 & -200.06 & -206.66 & -203.95 \\
CO$_{2}$ + 4H$_{2}$ $\rightarrow$ CH$_{4}$ + 2H$_{2}$O & -269.90 & -238.40 & -247.96 & -245.29 \\
CH$_{2}$O + 2H$_{2}$ $\rightarrow$ CH$_{4}$ + H$_{2}$O & -247.05 & -248.63 & -251.76 & -251.95 \\
CO + 3H$_{2}$ $\rightarrow$ CH$_{4}$ + H$_{2}$O & -277.51 & -269.90 & -275.87 & -273.80 \\
HCN + 3H$_{2}$ $\rightarrow$ CH$_{4}$ + NH$_{3}$ & -325.82 & -321.62 & -323.24 & -320.35 \\
H$_{2}$O$_{2}$ + H$_{2}$ $\rightarrow$ 2H$_{2}$O & -389.88 & -372.55 & -371.12 & -365.63 \\
HNO + 2H$_{2}$ $\rightarrow$ NH$_{3}$ + H$_{2}$O & -452.90 & -445.02 & -447.22 & -445.59 \\
C$_{2}$H$_{2}$ + 3H$_{2}$ $\rightarrow$ 2CH$_{4}$ & -451.58 & -441.34 & -450.02 & -446.71 \\
CH$_{2}$ + H$_{2}$ $\rightarrow$ CH$_{4}$ & -541.11 & -541.11 & -542.78 & -544.23 \\
F$_{2}$ + H$_{2}$ $\rightarrow$ 2HF & -572.62 & -559.49 & -568.01 & -564.93 \\
2CH$_{2}$ $\rightarrow$ C$_{2}$H$_{4}$ & -832.01 & -840.95 & -842.20 & -845.71 \\
\hline
Statistical measure & & & & \\
\hline
\hline
$\Delta_{max}^{abs}$ & 24.25 & 6.89 & 5.49 & \\
$\bar{\Delta}_{abs}$ & 8.65 & 3.84 & 2.51 & \\
$\Delta_{std}$ & 10.59 & 3.62 & 2.45 & \\
\hline
\end{tabular}
\caption{Reaction enthalpies and statistical measures of errors of 13 isogyric reactions (in kJ/mol), calculated with different
methods.}
\label{bra:tab2}
\end{center}
\end{table}

\begin{table}[h]
\begin{center}
\begin{tabular}{|c|c|c|c|c|c|}
\hline
Molecule & $N_{1}$ & $N_{2}$ & Molecule & $N_{1}$ & $N_{2}$ \\
\hline
H$_{2}$ & 500 & & C$_{2}$H$_{2}$ & 1000 & 1500 \\
HF & 750 & & HCN & 1000 & 1500\\
CH$_{2}$ & 1000 & & HNO & 1500 & 2000 \\
H$_{2}$O & 1000 & & F$_{2}$ & 1500 & 2000 \\
NH$_{3}$ & 1250 & & H$_{2}$O$_{2}$ & 1500 & 2000 \\
CH$_{4}$ & 1500 & & C$_{2}$H$_{4}$ & 2000 & 3000 \\
N$_{2}$ & 1000 & 1500 & CH$_{2}$O & 1500 & 2500 \\
CO & 1000 & 1500 & CO$_{2}$ & 2000 & 3000 \\
\hline
\end{tabular}
\caption{Number of points $N_{1}$ and $N_{2}$, used for extrapolation.}
\label{bra:tab3}
\end{center}
\end{table}
\newpage

\section*{Appendix A. Pseudocode.}\label{sec:Pseudocode}
\begin{itemize}
\setlength{\parskip}{0.001cm}
\small{
\item{Given: batch of samples of electronic coordinates $ \{ r_{1}^{\uparrow}...r_{N}^{\uparrow}; 
r_{1}^{\downarrow}...r_{N}^{\downarrow} \} $}
\item{Given: Nuclear coordinates $\{ R_{\alpha} \}$ }
\item{Construct: $f_{i\alpha}$, $f_{ij}$ $\forall \alpha,i,j$}
\item{\# ------ Preaparation of input layer features ------}
\item{\textbf{for each} $i$, spin \textbf{do}:  \hspace{1cm} \# Loop over spin up and spin down electrons $i$}
\item{\hspace{1cm} $\mathbf{h}^{input,spin}_{i}$ = concatenate($r_{i}^{spin}-R_{\alpha}$,$f_{i\alpha}$ $\forall \alpha$)}
\item{\hspace{1cm} \textbf{for each} $j$, spin1 \textbf{do}:}
\item{\hspace{2cm} $\mathbf{h}^{input,spin \hspace{0.1cm} spin1 }_{ij}$ 
= concatenate($r_{i}^{spin}-r_{j}^{spin1}$, $f_{ij}$ )}
\item{\# ------ Do this for all layers, except the last one ------}
\item{\textbf{for each} $l=0,\dots,L-1$ \textbf{do}:}
\item{\hspace{1cm} $\mathbf{g}^{l,\uparrow} = \frac{1}{n^{\uparrow}}\sum_{i}^{n^{\uparrow}}h_{i}^{l,\uparrow}$ }
\item{\hspace{1cm} $\mathbf{g}^{l,\downarrow} = \frac{1}{n^{\downarrow}}\sum_{i}^{n^{\downarrow}}h_{i}^{l,\downarrow}$ }
\item{\hspace{1cm} \textbf{for each} $i$, spin \textbf{do}:}
\item{\hspace{2cm} $\mathbf{g}^{l,spin,\uparrow}_{i} 
= \frac{1}{n^{\uparrow}}\sum_{j}^{n^{\uparrow}}h_{ij}^{l,spin,\uparrow}$ }
\item{\hspace{2cm} $\mathbf{g}^{l,spin,\downarrow}_{i}
= \frac{1}{n^{\uparrow}}\sum_{j}^{n^{\downarrow}}h_{ij}^{l,spin,\downarrow}$ }
\item{\hspace{2cm} $\mathbf{f}^{l,spin}_{i}$ = concatenate($\mathbf{h}^{l,spin}_{i}$,$\mathbf{g}^{l,\uparrow}$, 
$\mathbf{g}^{l,\downarrow}$, $\mathbf{g}^{l,spin,\uparrow}_{i}$, $\mathbf{g}^{l,spin,\downarrow}_{i}$ )}
\item{\hspace{2cm} $\mathbf{h}^{l+1,spin}_{i}$ = tanh(Conv1D($\mathbf{f}^{l,spin}_{i}$,$\mathbf{V}^{l})+\mathbf{b}^{l}$) 
+ $\mathbf{h}^{l,spin}_{i}$}
\item{\hspace{2cm} \textbf{for each} $j$, spin1 \textbf{do}:}
\item{\hspace{3cm} $\mathbf{h}^{l+1,spin,spin1}_{ij}$ = tanh(Conv2D($\mathbf{h}^{l,spin,spin1}_{ij}$,$\mathbf{W}^{l})+\mathbf{c}^{l}$) + $\mathbf{h}^{l,spin,spin1}_{ij}$}
\item{\# ------ Construct orbitals and wave function ------}
\item{\textbf{for each} determinant $k$ \textbf{do}:}
\item{\hspace{1cm} \textbf{for each} orbital $i$ \textbf{do}:}
\item{\hspace{2cm} \textbf{for each} $j$, spin \textbf{do}:}
\item{\hspace{3cm} e = envelope($r_{j}^{spin}$)}
\item{\hspace{3cm} $\varphi_{i}^{k,spin}(r_{j}^{k,spin}; \{ r_{/j}^{k,spin} \}; \{ r^{k,\overline{spin}}  \} )$ = $\big($
dot($\mathbf{w}_{i}^{k,spin}$,$\mathbf{h}^{L,spin}_{j}$)+$\mathbf{g}^{k,spin}_{i}$} $\big) \cdot e$
\item{\hspace{1cm} $D^{k\uparrow} = \det[ 
\varphi_{i}^{k,\uparrow}(r_{j}^{k,\uparrow}; \{ r_{/j}^{k,\uparrow} \}; \{ r^{k,\downarrow}  \} ]$ }
\item{\hspace{1cm} $D^{k\downarrow} = \det[
\varphi_{i}^{k,\downarrow}(r_{j}^{k,\downarrow}; \{ r_{/j}^{k,\downarrow} \}; \{ r^{k,\uparrow}  \} ]$ }
\item{$\Psi = \sum_{k} \omega_{k} D^{k\uparrow} D^{k\downarrow}$}}
\end{itemize}

\newpage
\begin{figure}[h!]
\begin{center}
\includegraphics[width=10cm,height=6cm]{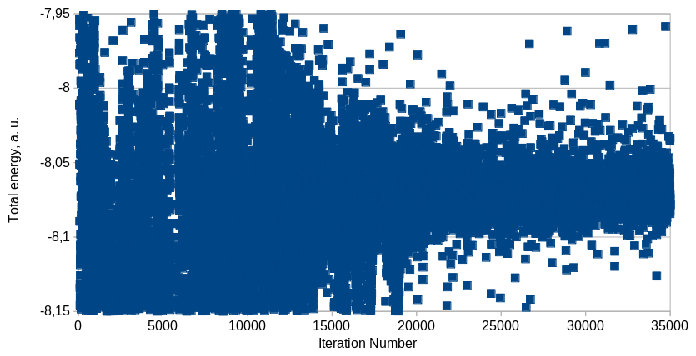}
\includegraphics[width=10cm,height=6cm]{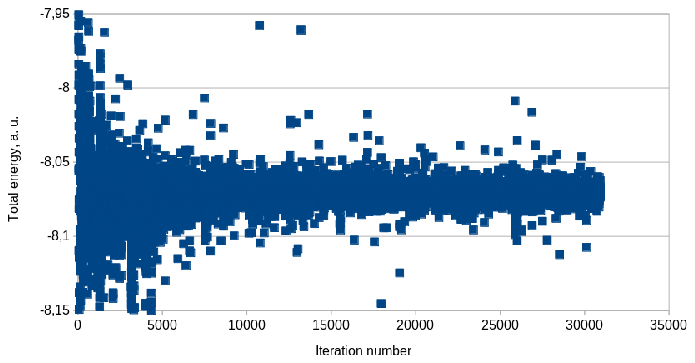}
\caption{Learning curves of LiH molecule, which are obtained with standard linear (upper pannel) an Slater exponential (proposed in this work, lower pannel) input features.}
\label{fig:Fig2}
\end{center}
\end{figure}

\begin{figure}[h!]
\begin{center}
\includegraphics[width=10cm,height=6cm]{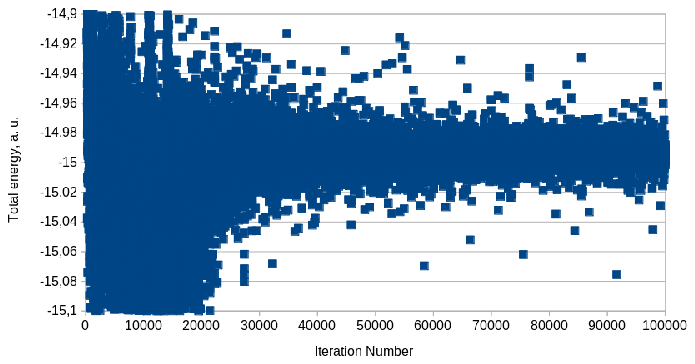}
\includegraphics[width=10cm,height=6cm]{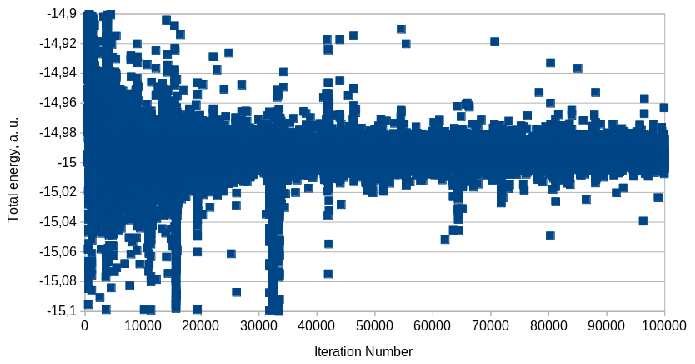}
\caption{Learning curves of Li$_{2}$ molecule, which are obtained with standard linear (upper pannel) and  Slater exponential (proposed in this work, lower pannel) input features.}
\label{fig:Fig3}
\end{center}
\end{figure}

\end{widetext}

\end{document}